\documentclass[10pt,a4paper]{article}
\usepackage{graphicx}
\usepackage{rotating}
\usepackage{amsmath,amssymb}
\begin{document}

\begin{center}
{\Large \bf
 The influence of the Alfv\'enic drift on the shape of cosmic ray spectra in SNRs}\\
\vspace{5mm}
 {\large \bf V.N.Zirakashvili, V.S.Ptuskin}
\end{center}

\noindent Pushkov Institite of Terrestrial Magnetism, Ionosphere and Radiowave \\
Propagation, Russian Academy of Sciences, 142190 Troitsk, Moscow Region, Russia

\begin{center}
{\large \bf Abstract} Cosmic ray acceleration in SNRs in the
presence of the Alfv\'enic drift is considered. It is shown that
spectra of accelerated particles may be considerably softer in the
presence of amplified  magnetic fields.

\end{center}

{\bf Introduction }

It is almost doubtless now that supernova remnants (SNRs) are the main source of galactic cosmic
rays (CR). The outer shell of the exploding star moves with a supersonic velocity and produces
a strong shock wave in the circumstellar medium. Diffusive shock acceleration
\cite{krymsky77, bell78} results in
the energy gain of energetic particles. The observation of high-energy TeV gamma-rays from
several SNRs is the evidence of effective acceleration of cosmic ray particles up to the
energy about 100 TeV
 \cite{aharonian07} in these objects. This energy is only a factor of 30 lower than the
"knee" energy $E_{knee}\sim $3 PeV in the CR spectrum.

At present there exist two numerical models of nonlinear diffusive
shock acceleration at the moving spherical shock of Berezhko and
co-authors \cite{berezhko96}, and Jones and co-authors
\cite{kang06}. A so-called shock modification by the CR pressure
is taken into account in these models. This is important since
about 10 percent of supernova energy is transferred to accelerated
CRs, if SNRs are the main source of CRs in the Galaxy. In addition
it seems that CR particles are accelerated only at some part of
the SNR shock, as it is observed in several SNRs.

These models may be used for modeling of CR acceleration in particular SNRs and for the calculation of an
overall  CR spectrum that is produced during lifetime of a SNR. It is expected that this spectrum is
close to   $E^{-2}$ (see \cite{ptuskin05}). Slightly harder spectrum
 $E^{-1.9}$ was numerically obtained recently \cite{berezhko07}.

In this short report we present results of the new numerical model of CR acceleration in SNRs. It is
shown, that CR spectra may be significantly softer if the advective velocity of CRs downstream of the shock is
essentially different from the gas velocity. This situation is probable if the magnetic field  is amplified in SNRs.
\\

{\bf Model of nonlinear diffusive shock acceleration in SNRs.}

Hydrodynamical equations for the gas density  $\rho (r,t)$, gas velocity $u(r,t)$, gas pressure
$P_g(r,t)$, and the equation for isotropic part of the CR momentum distribution
 $N(r,t,p)$ in the spherically symmetric case are given by

\begin{equation}
\frac {\partial \rho }{\partial t}=-\frac {1}{r^2}\frac {\partial }{\partial r}r^2u\rho
\end{equation}

\begin{equation}
\frac {\partial u}{\partial t}=-u\frac {\partial u}{\partial r}-\frac {1}{\rho }
\left( \frac {\partial P_g}{\partial r}+\frac {\partial P_c}{\partial r}\right)
\end{equation}

\begin{equation}
\frac {\partial P_g}{\partial t}=-u\frac {\partial P_g}{\partial r}
-\frac {\gamma _gP_g}{r^2}\frac {\partial r^2u}{\partial r}
-(\gamma _g-1)(w-u)\frac {\partial P_c}{\partial r}
\end{equation}

\begin{equation}
\frac {\partial N}{\partial t}=\frac {1}{r^2}\frac {\partial }{\partial r}r^2D(p,r,t)
\frac {\partial N}{\partial r}
-w\frac {\partial N}{\partial r}+\frac {\partial N}{\partial p}
\frac {p}{3r^2}\frac {\partial r^2w}{\partial r}
+\frac {\eta \delta (p-p_{inj})}{4\pi p^2_{inj}m}\rho (R+0,t)(\dot{R}-u(R+0,t))\delta (r-R(t))
\end{equation}
Here $P_c=4\pi \int p^2dpvpN/3$ is the CR pressure, $w(r,t)$ is the  advective velocity of CRs,
$\gamma _g$ is the adiabatic index of the gas, and $D(r,t,p)$ is the CR diffusion coefficient.
It was assumed that diffusive streaming of CRs results in the generation of magnetohydrodynamic (MHD)
waves. CR particles are scattered by these waves. That is why the CR advective velocity
 $w$ may differ from the gas velocity $u$. Damping of these waves results in additional gas heating. It is
described by the last term in Eq. (3). The last term in Eq. (4)
corresponds to the injection of thermal protons with momentum
$p=p_{inj}$ and mass $m$ at the shock front at $r=R(t)$. The
dimensionless parameter $\eta $ determines the injection
efficiency.

CR diffusion is determined by magnetic inhomogeneities. Strong
streaming of accelerated particles changes medium properties in
the shock vicinity. CR streaming instability results in the high
level of MHD turbulence  \cite{bell78} and even in the
amplification of magnetic field in young SNRs \cite{bell04}. Due
to this effect the maximum energy of accelerated particles may be
higher in comparison with previous estimates \cite{lagage83}.

According to the recent numerical modeling of this instability,
magnetic field is amplified by the flux of run-away highest energy
particles in the relatively broad region upstream of the shock
\cite{zirakashvili08}. Magnetic energy density is a small fraction
($\sim 10^{-3}$) of the energy density of accelerated particles.
This amplified almost isotropic magnetic field can be considered
as a large-scale magnetic field for lower energy particles which
are concentrated in the narrow region upstream of the shock.
Streaming instability of these particles produces MHD waves
propagating in the direction opposite to the CR gradient. This
gradient is negative upstream of the shock and MHD waves propagate
in the positive direction. The situation changes downstream of the
shock where CR gradient is as a rule positive and MHD waves
propagate in the negative direction. This effect is mostly pronounced
downstream of the shock because the magnetic field is additionally
amplified by the shock compression and the Alfv\'en velocity
 $V_A=B/\sqrt{4\pi \rho }$ may be comparable with the gas velocity in the shock frame
 $u'=\dot{R}-u(R-0,t)$. As for CR diffusion coefficient, it is probably close the Bohm value
$D_B=pvc/3qB$, where $q$ is the electric charge of
particles.
\\

{\bf Numerical modeling of CR acceleration in SNRs.} We apply
finite-difference method to solve Eqs (1-4) numerically upstream
and downstream of the shock. The auto-model variable  $\xi =r/R(t)$ is
used instead of radius $r$. The non-uniform numerical grid upstream of
the shock at
 $r>R$ allows to resolve small scales of hydrodynamical quantities appearing due to
the pressure gradient of low-energy CRs. Eq.
 (4) for CRs was solved using an implicit finite-difference scheme. The 
explicit conservative TVD scheme
\cite{trac03} for hydrodynamical equations (1-3) was used. These
equations are solved upstream the shock using the explicit
finite-difference scheme.

We shall assume that the coordinate dependencies of the magnetic
field and the gas density coincide:
\begin{equation}
B=\sqrt{4\pi \rho _0}\frac {\dot{R}\rho }{M_A\rho _0}
\end{equation}
Here $\rho _0$ is the gas density of the circumstellar medium. The parameter
  $M_A$ determines the value of the amplified magnetic field strength. The magnetic energy is about 3.5 percent
of the dynamical pressure
$\rho _0\dot{R}^2$, according to estimates from the width of X-ray filaments in young SNRs \cite{voelk05}. This
number and characteristic compression ratio of a modified SNR shock
 $\sigma =6$ correspond to  $M_A\approx 23$.

CR advective velocity differs from the gas velocity on the value of the radial component of the Alfv\'en velocity
$V_{Ar}$ calculated in the isotropic random magnetic field:  $w=u\pm V_{Ar}$. Here signs
$\pm $ correspond to regions upstream and downstream of the shock respectively. Using Eq. (5) we obtain
\begin{equation}
w=u\pm \frac {\dot{R}}{M_A}\sqrt{\frac {\rho }{3\rho _0}}
\end{equation}

We shall use CR diffusion coefficient $D=D_B$ calculated with the
magnetic field strength (5). Though in the real situation the
level of MHD turbulence may drop with distance upstream of the
shock
 and diffusion may be faster than the Bohm one (see \cite{zirakashvili08}),
we shall use these assumptions here (see also \cite{berezhko07}).

The numerical results are obtained for the SNR shock propagating in the medium with the 
number density
 $n_0=0.1$ cm$^{-3}$ and the temperature $T=10^4$ K. We use the ejecta mass
 $M_{ej}=1.4M_{\odot }$,  the energy of explosion $E_{SN}=10^{51}$ erg and the parameter of ejecta
velocity distribution $k=7$, corresponding to Ia supernovae. The initial shock velocity is
 $V_0=$31000 km/s. The injection efficiency is taken in the form $\eta =0.01\dot{R}/V_0$, and the injection
momentum is $p_{inj}=2m(\dot{R}-u(R+0,t))$. This dependence of the injection efficiency on the shock velocity
results in the significant shock modification already at early stages of SNR expansion. This is in agreement
with the observations of young extragalactic SNRs
 \cite{chevalier06} and with the modeling of collisionless shocks \cite{zirakashvili07}.

\begin{figure}[bt]
\includegraphics[height=14pc]{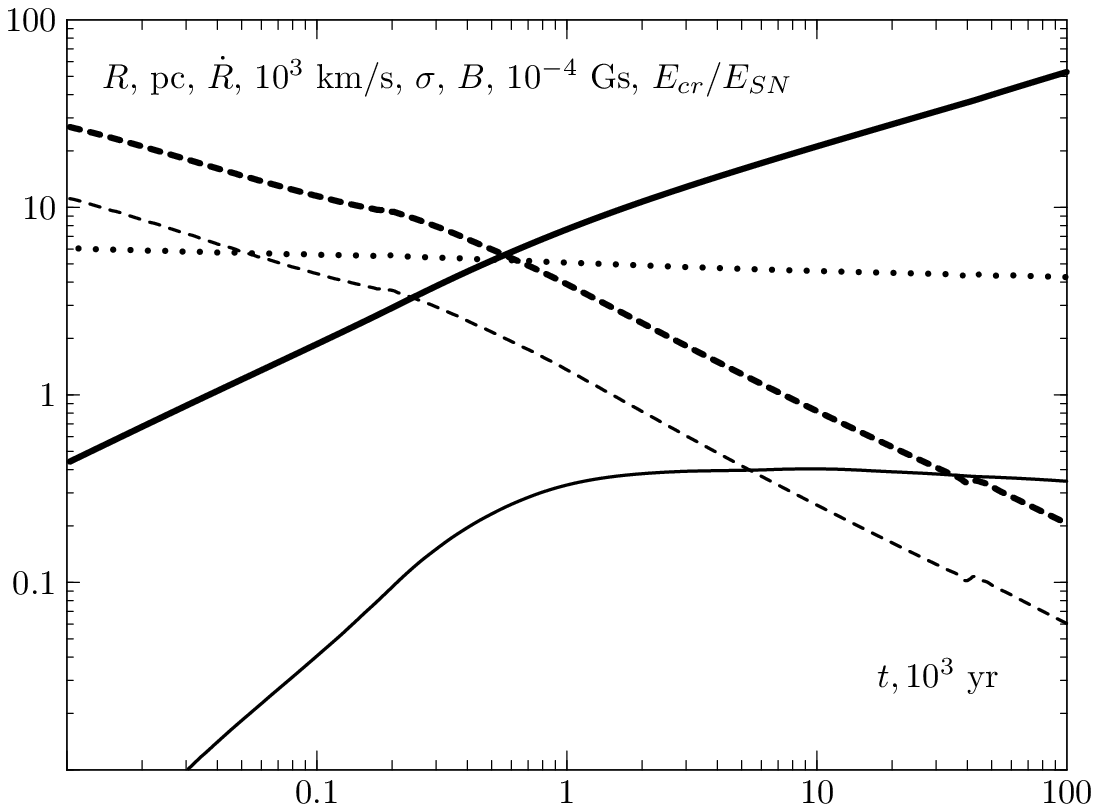}
\parbox[b]{7cm}{Fig.1. Dependencies on time of the shock radius $R$ (thick solid line), shock velocity $\dot{R}$
 (thick dashed line), the total compression ratio of the shock $\sigma $ (dotted line). Dependencies of the
magnetic field strength downstream of the shock (dashed line) and ratio of CR energy and energy of supernova
explosion  $E_{cr}/E_{SN}$ (solid line) are also shown. }
\end{figure}

The numerical results are obtained using the uniform grid with 800 cells downstream of the shock
for the variable  $\xi $, uniform grid with 800 cells upstream of the shock
for the variable $\ln (\xi -1+10^{-11})$.
The uniform grid with 200 cells for the variable $\ln p/mc$ is used.

The dependencies on time of the shock radius $R$, the shock
velocity $\dot{R}$, the total compression ratio of the shock
$\sigma $, the magnetic field strength downstream of the shock and
CR energy
 $E_{cr}/E_{SN}$ are shown in Fig.1. The calculations are performed until the beginning of the radiative phase of
SNR expansion at
 $t=10^5$ yr, when the value of the shock velocity drops down to $\dot{R}=206$ km/s. At this moment
of time the maximum energy of particles accelerated in SNR is
about 10 TeV, while higher energy particles have already leaved
the remnant. This maximum energy may be significantly lower if one
takes into account the wave damping on neutrals or nonlinear
damping \cite{ptuskin03}.

\begin{figure}[bt]
\includegraphics[height=14pc]{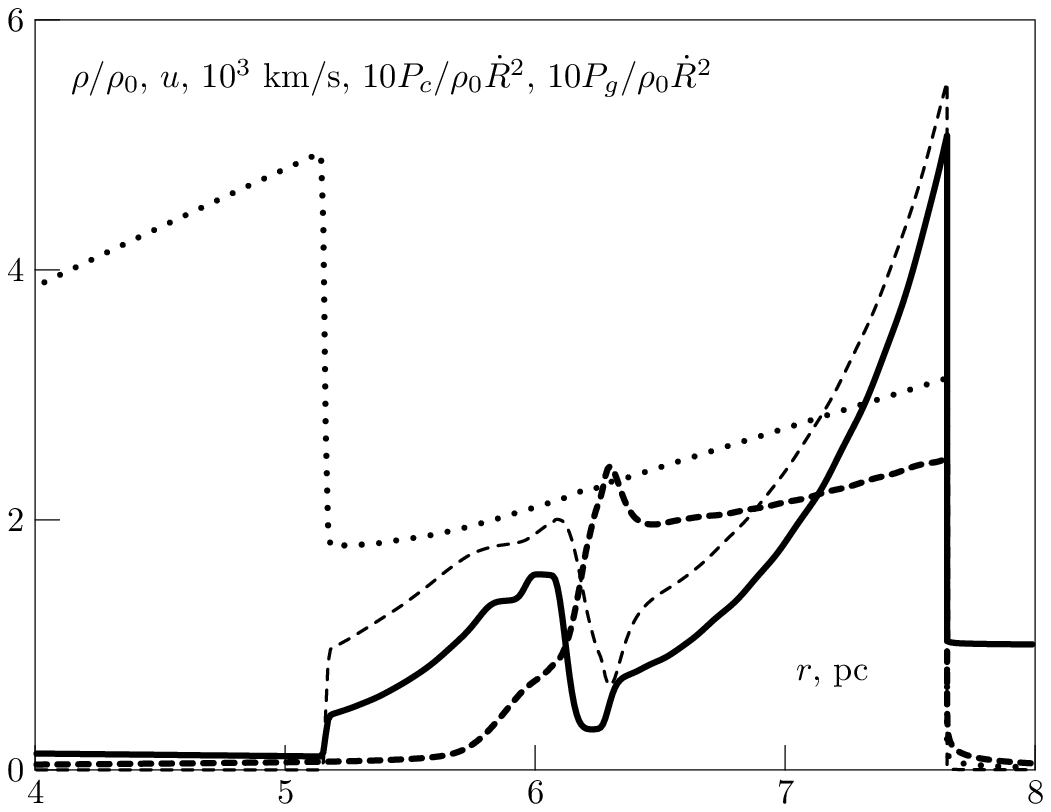}
\parbox[b]{7cm}{Fig.2. Radial dependencies of the gas density (thick solid line), the gas
velocity (dotted line), CR pressure (thick dashed line), the gas pressure (dashed line) at
 $t=10^3$ yr. At this moment of time the shock velocity is 3890 km/s, its radius is 7.6 pc, the magnetic
field strength downstream of the shock is 136 $\mu $G. }
\end{figure}

Radial dependencies of physical quantities at
 $t=10^3$ yr are shown in Fig.2. The contact discontinuity between the ejecta and the interstellar gas
is at
 $r=6.2$ pc. The reverse shock in the ejecta is situated at $r=5.2$ pc. We neglect the injection of
thermal ions into diffusive shock acceleration at the reverse shock. At the Sedov stage the reverse
shock moves in the negative direction and reach the center. The appearing reflected shock wave moves
in the positive direction and overtakes the main shock wave after 40 thousand years after explosion. This
results in the non-monotonic behavior of the shock velocity at this moment of time in Fig.1.

The spectrum of CR protons at the shock at $t=10^3$ yr is shown in
Fig.3 (solid line). It is compared with the proton spectrum
obtained when the CR advection velocity coincides with the gas
velocity downstream of the shock (dashed line). CR pressure is
$P_c=0.25\rho _0\dot{R}^2$ in the first case and a factor of two
higher in the second one. The CR spectrum is significantly softer
in the first case.

\begin{figure}[bt]
\includegraphics[height=14pc]{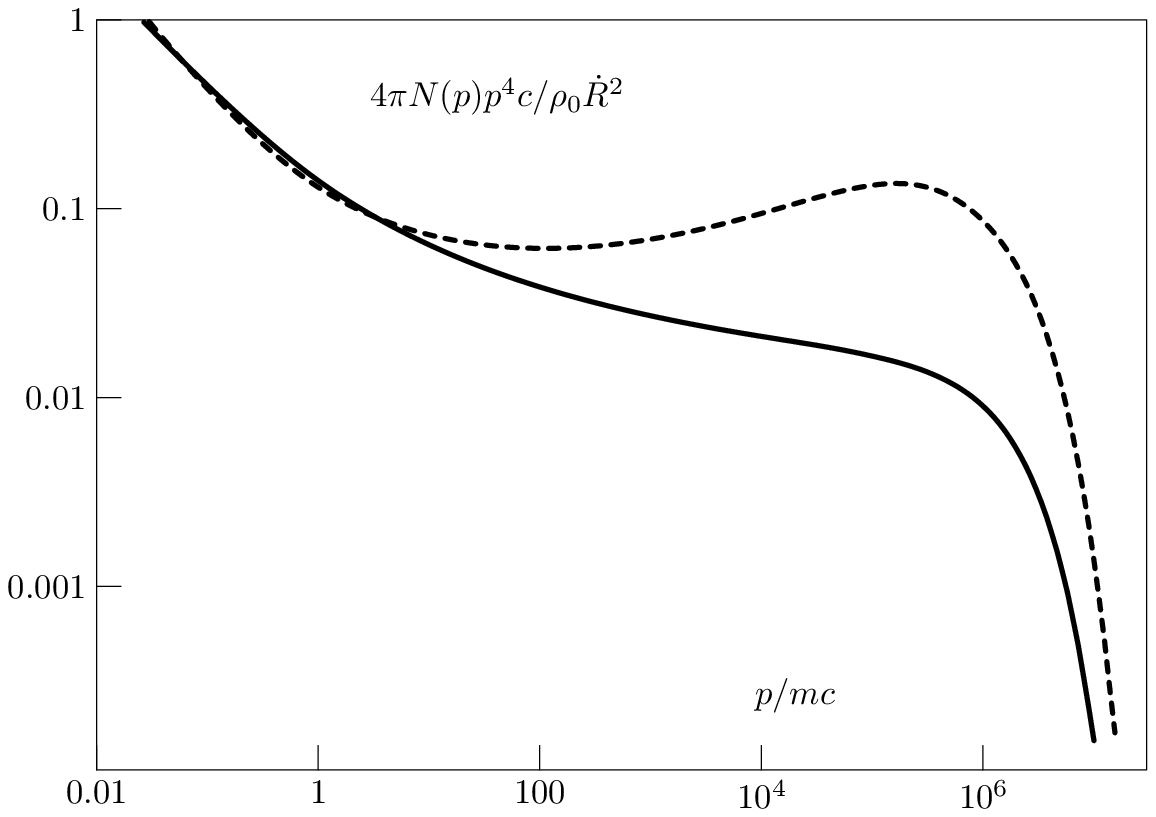}
\parbox[b]{7cm}{Fig.3. Spectrum of CR protons at the shock front at $10^3$ yr after
explosion. The results corresponding to the Eq. (6) (solid line)
and to CR advection velocity coinciding with the gas velocity
downstream of the shock (dashed line) are shown.}
\end{figure}

The spectrum of CR protons produced during the lifetime of the SNR
 $F(p)$ is shown in Fig.(4) (solid line). It is determined by the integration of
CR momentum distribution $N(p)$ on the simulation volume at the
moment $t=10^5$ yr. It is compared  with the CR proton spectrum
obtained when the CR advection velocity coincides with the gas
velocity downstream of the shock (dashed line). The last curve is
similar to results of Berezhko and V\"olk \cite{berezhko07}. The
integrated CR spectrum is again significantly softer in the first
case.
\\

\begin{figure}[bt]
\includegraphics[height=14pc]{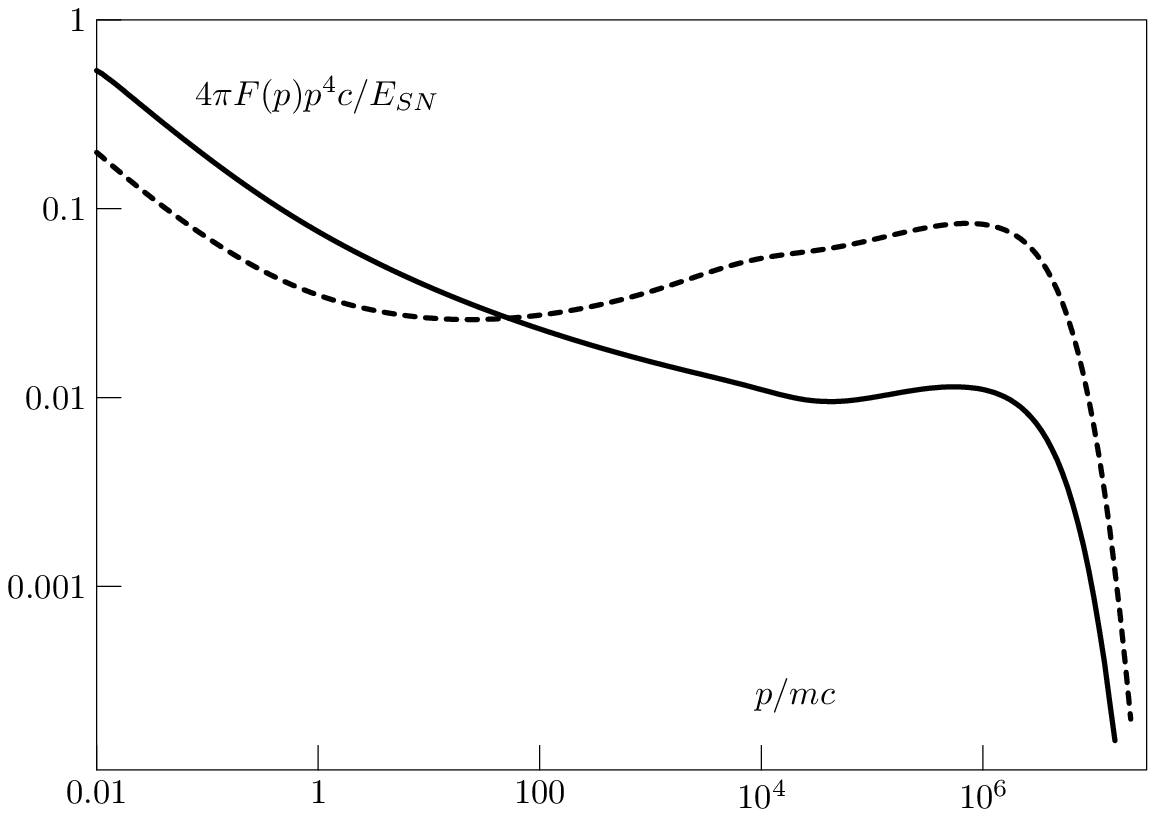}
\parbox[b]{7cm}{Fig.4. Spectrum of CR protons produced during the lifetime of SNR. 
The results corresponding to the
Eq. (6) (solid line) and to CR advection velocity coinciding with
the gas velocity downstream of the shock (dashed line) are shown.}
\end{figure}

{\bf Conclusion. } Magnetic field amplification in SNRs may result
in the significant difference of the CR advective velocity and the gas
velocity downstream of the SNR shock. CR spectra in young SNRs and
CR spectra produced by SNR during its lifetime may be
significantly softer due to this effect. The fraction of supernova
energy transferred to CRs is also reduced from 75 percent down to
35 percent of the mechanical energy of explosion.
SNR luminosity in the TeV gamma-rays is reduced also. \\
This work was supported by RFBR 07-02-00028 grant.

\end{document}